\documentclass[pre,twocolumn,showpacs]{revtex4-1}

\usepackage[dvips]{color}

\usepackage{graphicx}
\usepackage{amsmath}

\begin{document}

\title{The double-layer of penetrable ions:  an alternative route to charge reversal}

\author{Derek Frydel}
\affiliation{Institute of Physics, the Federal University of Rio Grande do Sul, 
PO Box 15051, 91501-970, Porto Alegre, RS, Brazil}
\author{Yan Levin}
\affiliation{Institute of Physics, the Federal University of Rio Grande do Sul, 
PO Box 15051, 91501-970, Porto Alegre, RS, Brazil}

\date{\today}

\begin{abstract}
We investigate a double-layer of penetrable ions near a charged wall.  We find a new mechanism 
for charge reversal that occurs in the weak-coupling regime and, accordingly, 
the system is suitable for the mean-field analysis.  The penetrability is achieved by smearing-out 
the ionic charge inside a sphere, so there is no need to introduce non-electrostatic forces and 
the system in the low coupling limit can be described by a modified version of the 
Poisson-Boltzmann equation.  The predictions of the theory are compared with the Monte 
Carlo simulations. 
\end{abstract}

\pacs{
}

\maketitle

\section{Introduction}

Overcharging is a surprising phenomenon in which conterions adsorbed to a surface exceed the 
number of fixed surface charges~\cite{Pa80}.  As a consequence, coions are drawn from the 
bulk toward the overcharged surface leading to a formation of a "triple-layer".  Because the 
underlying mechanism of overcharging relies on electrostatic correlations~\cite{Reviews}, 
it has become synonymous 
with the strong-coupling limit.  Indeed, a mean-field treatment which accurately captures the 
weak-coupling limit, cannot describe overcharging.  (However, the mean-field in combination with 
the excluded volume interactions can induce overcharging if the bulk volume fraction of an 
electrolyte is sufficiently large to generate a depletion force that pushes particles against a surface.   This effect is seen in uncharged systems and persists for weakly charged surface 
charges \cite{Kjellander98,Lozada02,Lozada10,Frydel12}.)

In the strong-coupling limit overcharging is the result of increased structuring within the layer of 
counterions.  The electrostatic correlations between the condensed counterions  lead  to 
formation of correlation "holes" within the layer of condensed ions which can attract excess of 
counterions from the bulk.  The value of the coupling constant 
$\Gamma=Z^3\lambda_B/\lambda_{\rm GC}$, which is the ratio between the Bjerrum and 
Gouy-Chapman length ($Z$ is the valency), estimates the extent of correlation effects.  In the 
limit $\Gamma\to\infty$, the counterions are said to freeze into a 2d Wigner crystal 
\cite{Shklovskii99}.  At large, but finite $\Gamma$, the local structure of an ionic fluid remains 
Wigner-like \cite{Ro96,Shklovskii99,Sal05,Trizac11}.  The above mechanism is specific to 
Coulomb interactions that diverge as $r\to 0$ and, therefore, exhibit the excluded volume 
effects \cite{Huang93,Sal05}.  If, however, the divergence in the pair interaction is removed (the 
pair potential is bounded as $r\to 0$), particles can interpenetrate and the usual excluded 
volume interactions underlying the crystal formation are eliminated (at sufficiently high 
temperature and/or density).  For some family of bounded potentials, particles can form stacks 
where two or more particles occupy the same position and act as an effective single particle.  
This is possible only if a pair potential is sufficiently flat around $r=0$ \cite{Lowen01}.  One  
example is the penetrable sphere model where the pair potential is the step function 
\cite{Witten89,Lowen98,Schmidt99,Rosenfeld00}.  The stacking formations stabilize the liquid 
phase \cite{Lowen98}, since doublets, triplets, etc. effectively decrease the number of particles.  
The presence of stacked formations is signaled as a positive peak in the pair 
correlation function at $r\to 0$ \cite{Lowen98}.  Extrapolating these ideas to ions, which in 
addition to penetrable cores have long-range Coulomb interactions, we ask what influence
penetrability has on the structure of a double-layer.  Can the restructuring invoked by the 
penetrability lead to overcharging in the weak-coupling limit?

In the present work, the divergence in the Coulomb potential is removed by smearing-out the central charge of an ion over a finite region.  The penetrating core, then, depends on the weight
function used to smear out the charge.  This procedure does not require going beyond the 
framework of electrostatics and the weak-coupling limit can be described by the modified 
version of the Poisson-Boltzmann equation.

Bulk properties and phase diagram of penetrable ions have been investigated in  
\cite{Hansen11a,Hansen11b,Hansen12}.  The main feature is the formation of the Bjerrum pairs 
of two opposite ions that function as polarizable particles.  The formation of these pairs leads 
to an insulator-conductor transition~\cite{Hansen11a}, which does not exist in systems of 
hard-core ions in three dimensions \cite{Yan93}.  Thus, penetrability dramatically affects
the phase transition and the topology of the phase diagram.

The model of penetrable particles is not only of theoretical interest.  Various macroparticles can 
exhibit interpenetration.  Marquest and Witten \cite{Witten89} suggested a penetrable sphere 
model for micelles.   Polymer coils and dendrimers in good solvent can be represented by a 
Gaussian core model \cite{Hansen00,Schmidt01}.  If in addition these macroparticles are 
charged, as is often the case for real systems \cite{Barrat95}, then the model of penetrable ions 
can be of genuine physical relevance.  Recently, ionic microgels have been 
modeled as uniformly charged spheres \cite{Lowen04}, allowing interpenetration at short- and
Coulomb interactions at long-separations.

\section{The Poisson-Boltzmann equation for smeared-out charges}

The charge $q_i$ of an ion $i$ is smeared out over the finite region according to the weight 
function $\omega_i({\bf r}-{\bf r}_i)$ such that $\int d{\bf r}\,\omega_i({\bf r}-{\bf r}_i)=q_i$.  
The charge density operator for $N$ smeared-out ions is
\begin{equation}
\hat\rho_c({\bf r})=\sum_{i=1}^{N}\int d{\bf r}' \delta({\bf r}'-{\bf r}_i)\omega_i({\bf r}'-{\bf r}).
\end{equation}
In this work we consider a symmetric electrolyte, $q_+=-q_-$.  The averaged charge density 
for this system is
\begin{equation}
\rho_c({\bf r}) = \int d{\bf r}' \Bigg[\rho_+({\bf r}')\omega_+({\bf r}'-{\bf r})
+ \rho_-({\bf r}')\omega_-({\bf r}'-{\bf r})\Bigg],
\label{eq:rhoc}
\end{equation}
where $\rho_+$ and $\rho_-$ denotes the number density of anions and cations, respectively.  
The Poisson equation is 
\begin{equation}
\epsilon\nabla^2\psi = -\int d{\bf r}' \Bigg[\rho_+({\bf r}')\omega_+({\bf r}'-{\bf r})
+ \rho_-({\bf r}')\omega_-({\bf r}'-{\bf r})\Bigg],
\end{equation}
where $\epsilon$ is the background dielectric constant.  To obtain a closed equation, we need an 
expression for $\rho_-$ and $\rho_+$ in terms of the mean electrostatic potential $\psi$.  For 
point charges this leads to 
\begin{equation}
\rho_{\alpha}({\bf r}) = c_se^{-\beta q_{\alpha}\psi({\bf r})},
\end{equation}
where the subscript $\alpha$ is either $+$ or $-$, and $c_s$ is the bulk salt concentration.  The 
number density depends locally on an electrostatic potential.  However, if charge is smeared 
around the ion center at ${\bf r}$, the entire distribution $\omega({\bf r}'-{\bf r})$ interacts 
with the mean electrostatic potential, 
\begin{equation}
\psi_{\alpha}({\bf r}) = \int d{\bf r}'\psi({\bf r}')\omega_{\alpha}({\bf r}'-{\bf r}), 
\end{equation}
and the number density becomes
\begin{equation}
\rho_{\alpha} = c_s e^{-\beta\int d{\bf r}'\psi({\bf r}')\omega_{\alpha}({\bf r}'-{\bf r}) }.
\label{eq:rho_alpha}
\end{equation}
We may now write down the mean-field equation for the electrostatic potential produced by
smeared-out ions, 
\begin{eqnarray}
-\epsilon\nabla^2\psi &=& c_s\int d{\bf r}'\omega_+({\bf r}'-{\bf r})
e^{-\beta \int d{\bf r}''\psi({\bf r}'')\omega_+({\bf r}''-{\bf r}')}\nonumber\\
&+&c_s\int d{\bf r}'\omega_-({\bf r}'-{\bf r})
e^{-\beta \int d{\bf r}''\psi({\bf r}'')\omega_-({\bf r}''-{\bf r}')}.\nonumber\\
\label{eq:FSPB}
\end{eqnarray}
We refer to this modified Poisson-Boltzmann equation the Finite-Spread PB equation (FSPB).  
The FSPB equation complements the already quite 
sizable set of modified PB equations:  the PB that incorporates the excluded volume 
interactions \cite{david97}, the dipolar interactions \cite{david07}, the nonlinear solvent 
contributions \cite{Frydel11a}, and the polarizability of ions \cite{Frydel11,Roji12}.  
The idea of finitely spread-out ions was considered in \cite{Wang10} to study the self energy
contributions beyond the mean-field.
Mathematically, the FSPB equation has the structure of a convoluted equation.  Each ion is 
convoluted according to the weight function which determines the composition of a single ion.

The Eq.~(\ref{eq:FSPB}) can also be obtained by minimizing the grand potential,
\begin{eqnarray}
{\Omega}[\{\rho_{\alpha}\}] &=& 
\frac{1}{2}\int d{\bf r}\int d{\bf r}'\,
\frac{\rho_c({\bf r})\rho_c({\bf r}')}{4\pi\epsilon|{\bf r}'-{\bf r}|}
\nonumber\\
&+& k_BT \sum_{\alpha}\int d{\bf r}\,\rho_{\alpha}({\bf r})
\Big[\log\rho_{\alpha}\Lambda^3-1\Big]
\nonumber\\
&-&\sum_{\alpha}\mu\int d{\bf r}\,\rho_{\alpha}({\bf r}),
\label{eq:Omega}
\end{eqnarray}
where $\Lambda$ is the thermal de Broglie wavelength and $\mu=\mu_{+}=\mu_{-}$ is the 
chemical potential.  The three contributions are the electrostatic energy, entropy, and we allow
the number density to fluctuate through the contact with a reservoir.  The minimization 
$\frac{\delta{\Omega}}{\delta\rho_{\alpha}}=0$ recovers 
Eq. (\ref{eq:rho_alpha}) and the application of the Poisson equation yields the FSPB model.

\section{Various Distributions $\omega(r)$}
The concept of charge smearing is not novel to this work, but it has been evoked many times in 
the past both as a physical representation and a mathematical construct.   The best known 
example (and to our knowledge the earliest) of mathematical construct is the calculation of 
Ewald summation to treat periodic charges \cite{Ewald21}, today the most practiced method to 
account for contributions due to periodic boundary conditions of a simulated system 
\cite{Frenkel02}, where by spreading the charge one achieves separation of interactions into 
the short- and long-range counterpart.  Another instance of mathematical convenience 
is the Onsager smearing optimization to obtain the variational free energy of the strongly 
correlated one component plasma \cite{Rosenfeld87}.  The idea of charges which at short
separations exhibit soft repulsion and at long separations the Coulomb interaction appeared
in \cite{Hansen81} to represent semi-classical hydrogen plasma at high temperature and 
density, yet no smearing procedure was used to construct this potential.  The actual smearing
procedure to represent physical particles was used to model electrons set in cyclotron motion by 
a uniform magnetic field \cite{Shklovskii96,Shklovskii96b}.  
In soft matter, the smearing out procedure was used to represent microgels in \cite{Denton03},
and recently to represent charged polymers \cite{Hansen11a,Hansen11b,Hansen12}.

We consider a few simple, spherically symmetric distributions $\omega(r)$ and 
the interactions that result from these distributions.  A spherically symmetric charge distribution 
generates the following electrostatic potential,
\begin{equation}
\psi(r) = \frac{1}{\epsilon}\Bigg[\frac{1}{r}\int_0^rds\,s^2\omega(s)
+ \int_r^{\infty}ds\,s\, \omega(s)\Bigg].
\end{equation}
To derive the above formula we applied the Gauss law to the Poisson equation, and afterwards 
we integrated the resulting electrostatic field (the integration by parts was evoked).   
If the distribution is uniform inside a spherical cavity,
\begin{equation}
\omega({\bf r}'-{\bf r}) = \frac{3Q}{4\pi R^3}\theta(R-|{\bf r}'-{\bf r}|), 
\label{eq:h_theta}
\end{equation}
where $\theta$ is the Heaviside function and $\int d{\bf r}'\omega({\bf r}'-{\bf r})=Q$,
then the potential inside a sphere is
\begin{equation}
\psi(r\le R) = \frac{Q}{4\pi\epsilon}\Bigg[\frac{3R^2-r^2}{2R^3}\Bigg].
\label{eq:psi_ps}
\end{equation}
Outside the sphere the Coulomb potential is recovered.  
The interaction between two ions with this charge distribution is \cite{Denton03,Rosenfeld87}
\begin{equation}
U_{\rm}(r\le 2R) = 
\frac{Q^2}{4\pi\epsilon}\Bigg[\frac{192R^5-80R^3r^2+30R^2r^3-r^5}{160 R^6}\Bigg].
\label{eq:U_ss}
\end{equation}
For non-overlapping separations, ions behave like point charges.  
Next, we consider the distribution 
\begin{equation}
\omega({\bf r}'-{\bf r}) = \frac{Q}{4\pi R^2}\delta(R-|{\bf r}'-{\bf r}|), 
\label{eq:h_delta}
\end{equation}
where a charge is smeared-out over a spherical shell.  Inside a sphere, the electrostatic 
potential is constant,
\begin{equation}
\psi(r\le R) = \frac{Q}{4\pi\epsilon}\frac{1}{R}.  
\end{equation}
The interaction between two of these distributions for overlapping separations is
\begin{equation}
U(r\le 2R) = \frac{Q^2}{4\pi\epsilon}\Bigg[\frac{4R-r}{4R^2}\Bigg].
\end{equation}
A sphere and a shell distributions interact via
\begin{equation}
U(r\le 2R) = \frac{Q^2}{4\pi\epsilon}\Bigg[\frac{16R^3-4Rr^2+r^3}{16R^4}\Bigg].
\end{equation}
Finally, we consider a Gaussian distribution \cite{Ewald21,Hansen11a,Hansen11b,Hansen12}, 
\begin{equation}
\omega({\bf r}'-{\bf r}) = \frac{Q^2}{(2\pi)^{3/2} R^3}e^{-r^2/2R^2},
\label{eq:h_G}
\end{equation}
which leads to the following pair interaction
\begin{equation}
U(r) = \frac{Q^2}{4\pi\epsilon}\frac{{\rm erf}(r/2R)}{r}.
\label{eq:U_G}
\end{equation}

\section{Numerical results for the FSPB equation}
We primarily focus on the distribution
\begin{equation}
\omega_{\pm}(|{\bf r}-{\bf r}_i|) = \pm e\theta(R-|{\bf r}-{\bf r}_i|)/\nu.
\label{eq:h1}
\end{equation}  
where $\nu=4\pi R^3/3$ is the ionic volume.  The FSPB equation in reduced units reads
\begin{eqnarray}
\nabla^2\phi &=& 
\kappa^2\nu^{-1}\int d{\bf r}'\theta(R-|{\bf r}'-{\bf r}|)\sinh(\bar\phi_{\theta}),\nonumber
\label{eq:FSPB_1}
\end{eqnarray}
where 
\begin{equation}
\bar\phi_{\theta}({\bf r}) = \frac{1}{\nu}\int d{\bf r}' \phi({\bf r}')\theta(R-|{\bf r}'-{\bf r}|),
\end{equation}
$\phi=\beta e \psi$, $\kappa^{-1}=1/\sqrt{8\pi\lambda_B c_s}$ is the screening length, and 
$\lambda_B=\beta e^2/4\pi\epsilon$ is the Bjerrum length.  
The number density of each species is
\begin{equation}
\rho_{\pm} = c_se^{\mp\bar\phi_{\theta}}.
\end{equation}
All the ions occupy the half-space $x>0$.  The charged hard wall at $x=-R$ determines  
boundary conditions,
\begin{equation}
\frac{\partial \phi}{\partial x} = -4\pi\lambda_B\sigma_c,
\end{equation}
where $\sigma_c$ is the surface charge.  Note that the number density is limited to the region 
$x>0$, but the charge density, because of the distribution $\omega(r)$, reaches all the way to 
$x=-R$, the location of the charged wall.  Consequently, we assume the (hard-sphere)-(hard-wall) 
potential between an ion and the wall.  Subsequent figures show data points for $x>0$, the 
region available to ion centers.   

In the dilute limit ($c_s\to 0$), overcharging does not happen, the solution of the PB equation 
for point ions in the weak-coupling limit yields an algebraic density profile,
\begin{equation}
\rho(x)=\frac{2\pi\lambda_B\sigma_c^2}{(1+x\lambda_{\rm GC}^{-1})^2}.
\label{eq:WCL}
\end{equation}
In the strong-coupling limit the electrostatic correlations modify the functional form of the
distribution \cite{Netz00},
\begin{equation}
\rho(x)=2\pi\lambda_B\sigma_c^2\exp(-x\lambda_{\rm GC}^{-1}),
\label{eq:SCL}
\end{equation}
where $\lambda_{\rm GC}=(2\pi\lambda_B\sigma_c)^{-1}$ is the Gouy-Chapman length.  The 
barometric-like distribution in the transverse direction is a consequence of high degree of 
ordering in the lateral direction reminiscent of the Wigner crystal \cite{Shklovskii99,Trizac11}.

\begin{figure}[tbh]
\vspace{0.6cm}
\centerline{\resizebox{0.45\textwidth}{!}
{\includegraphics{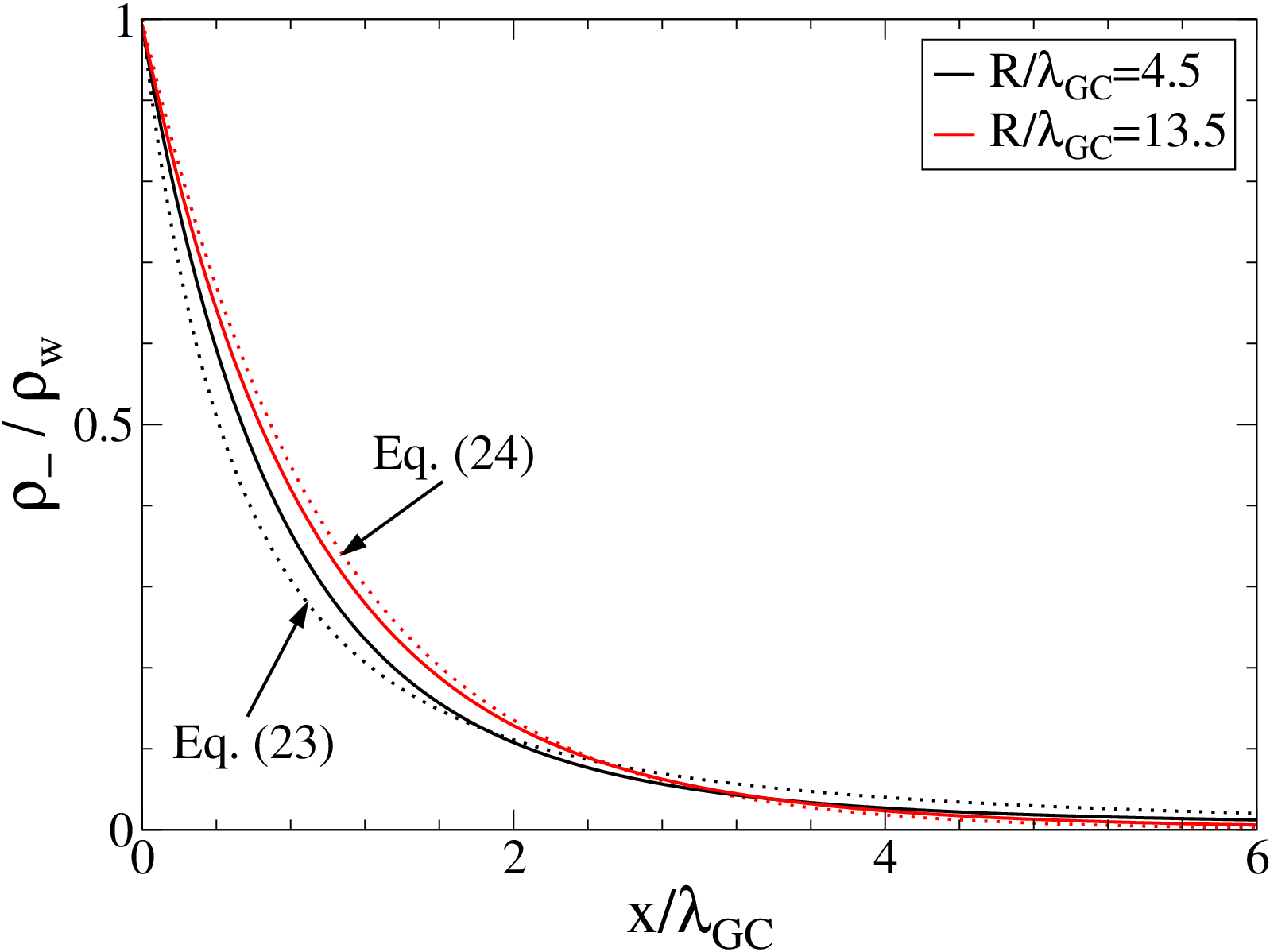}}}
\caption{The counterion density profile in the dilute limit (no coions).  The dotted lines demarcate 
the limiting behaviors in Eq. (\ref{eq:WCL}) and Eq.~(\ref{eq:SCL}).  The wall charge is positive 
and the counterions have negative charge.  As $R$ increases, the profile takes the exponential 
shape.  }
\label{fig:rho_counter_SCL2}
\end{figure}

Although the FSPB equation is purely mean-field, we find similar modification of the density 
profile, but this time the relevant parameter is $R$.  The limit $R\to 0$ corresponds to  the ionic 
distribution given by Eq.~(\ref{eq:WCL}), and the limit $R\to\infty$ to the one given by 
Eq.~(\ref{eq:SCL}), see Fig.~(\ref{fig:rho_counter_SCL2}).  In the limit $R\to\infty$, see
Eq.~(\ref{eq:U_ss}), we recover an ideal gas particles in a uniform 
gravitational field.  

\begin{figure}[tbh]
\vspace{0.6cm}
\centerline{\resizebox{0.45\textwidth}{!}
{\includegraphics{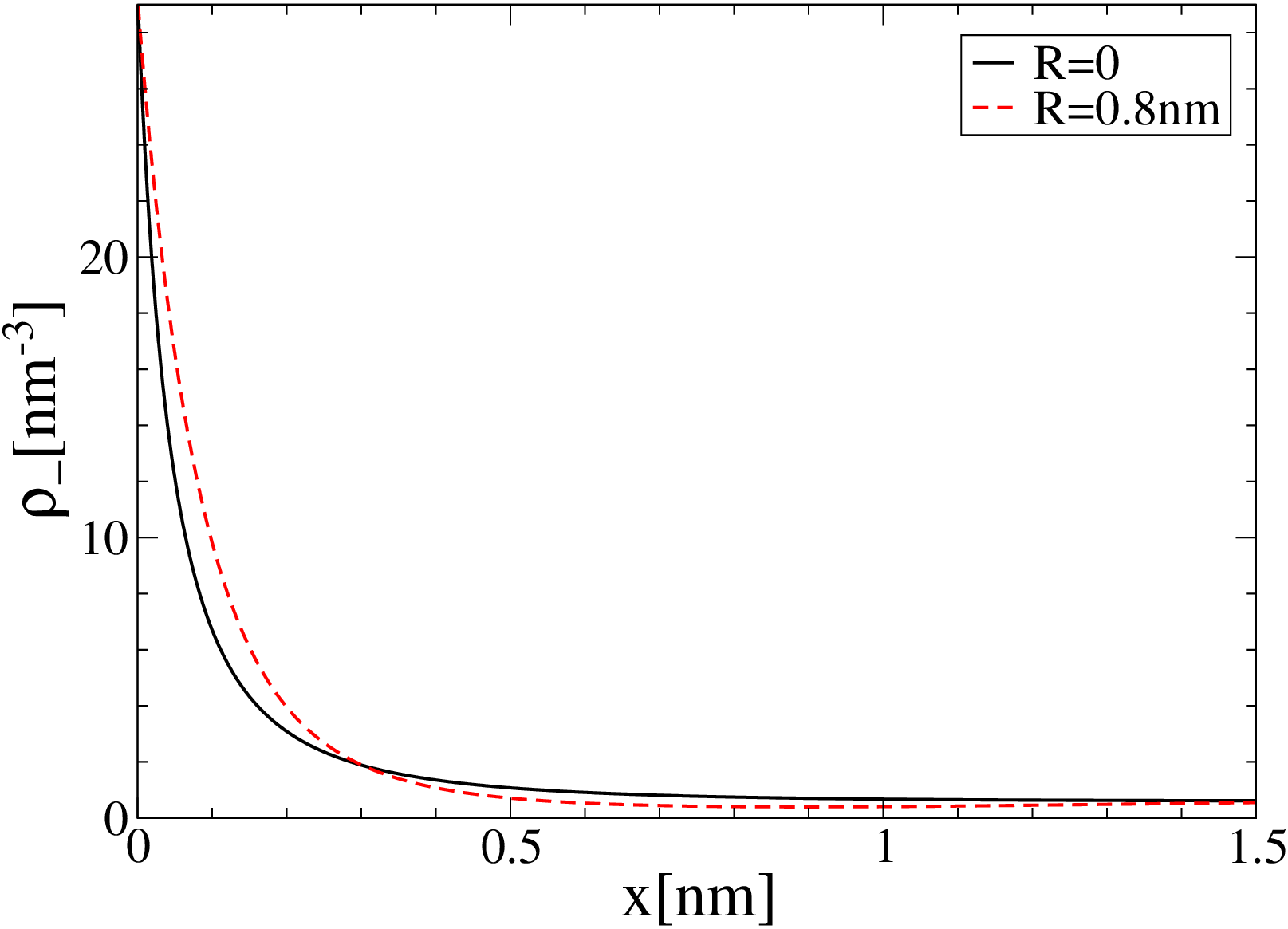}}}
\caption{The counterion density profile near a charged wall.  
$R=0$ corresponds to point ions.  The relevant lengths are:
$\kappa^{-1}=0.3{\rm nm}$ $\lambda_B=0.72{\rm nm}$, and 
$\lambda_{\rm GC}=0.09{\rm nm}$.}
\label{fig:rho_counter_R8_Z0_1.0M}
\end{figure}

\begin{figure}[tbh]
\vspace{0.6cm}
\centerline{\resizebox{0.45\textwidth}{!}
{\includegraphics{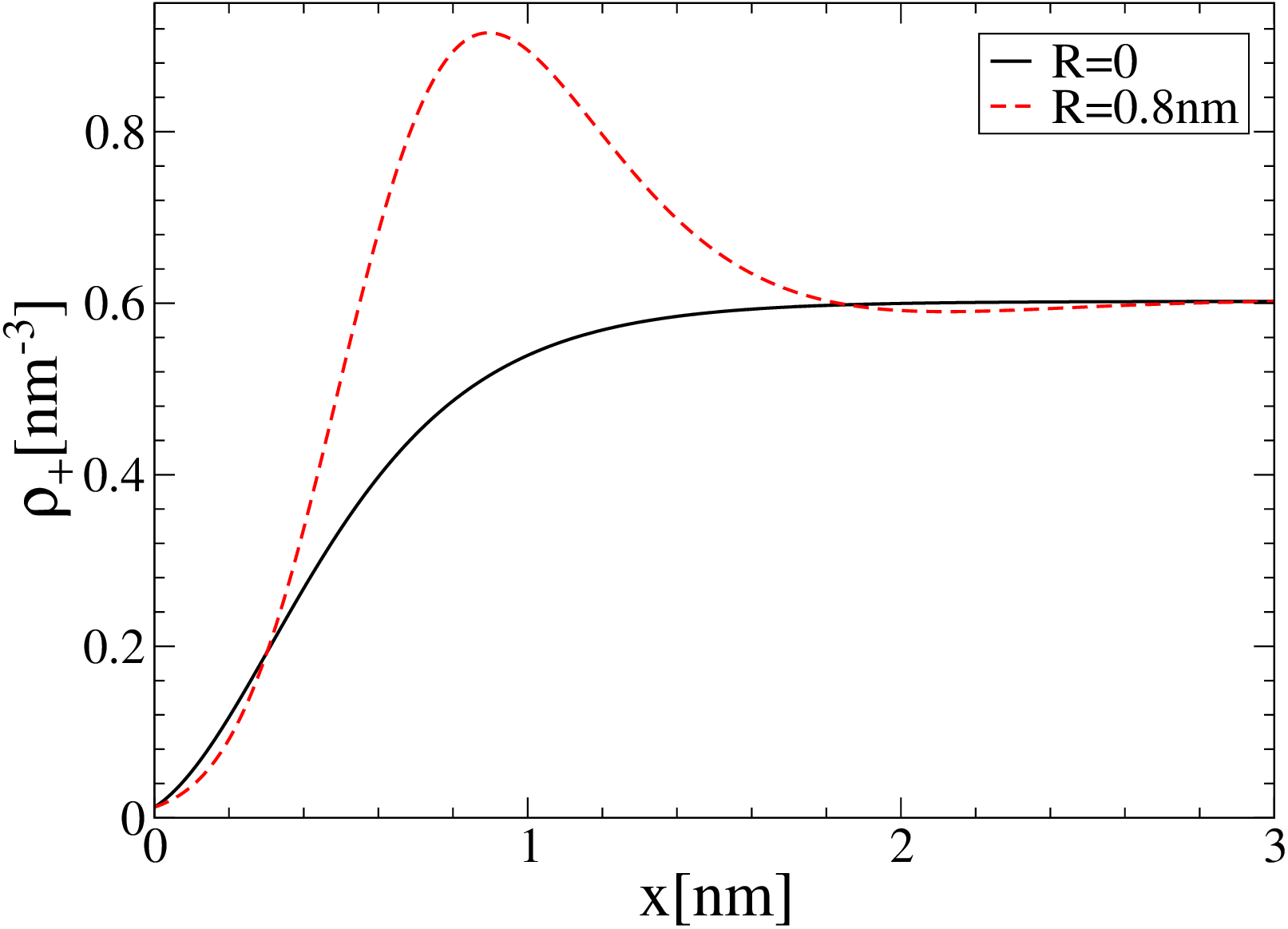}}}
\caption{The coion distribution at a charged wall corresponding to the system in
Fig.~(\ref{fig:rho_counter_R8_Z0_1.0M}).  The coions are in excess to neutralize the inverted 
charge.  Note that the density profile exhibits oscillations.  }
\label{fig:rho_co_R8_Z0_1.0M}
\end{figure}

\begin{figure}[tbh]
\vspace{0.6cm}
\centerline{\resizebox{0.45\textwidth}{!}
{\includegraphics{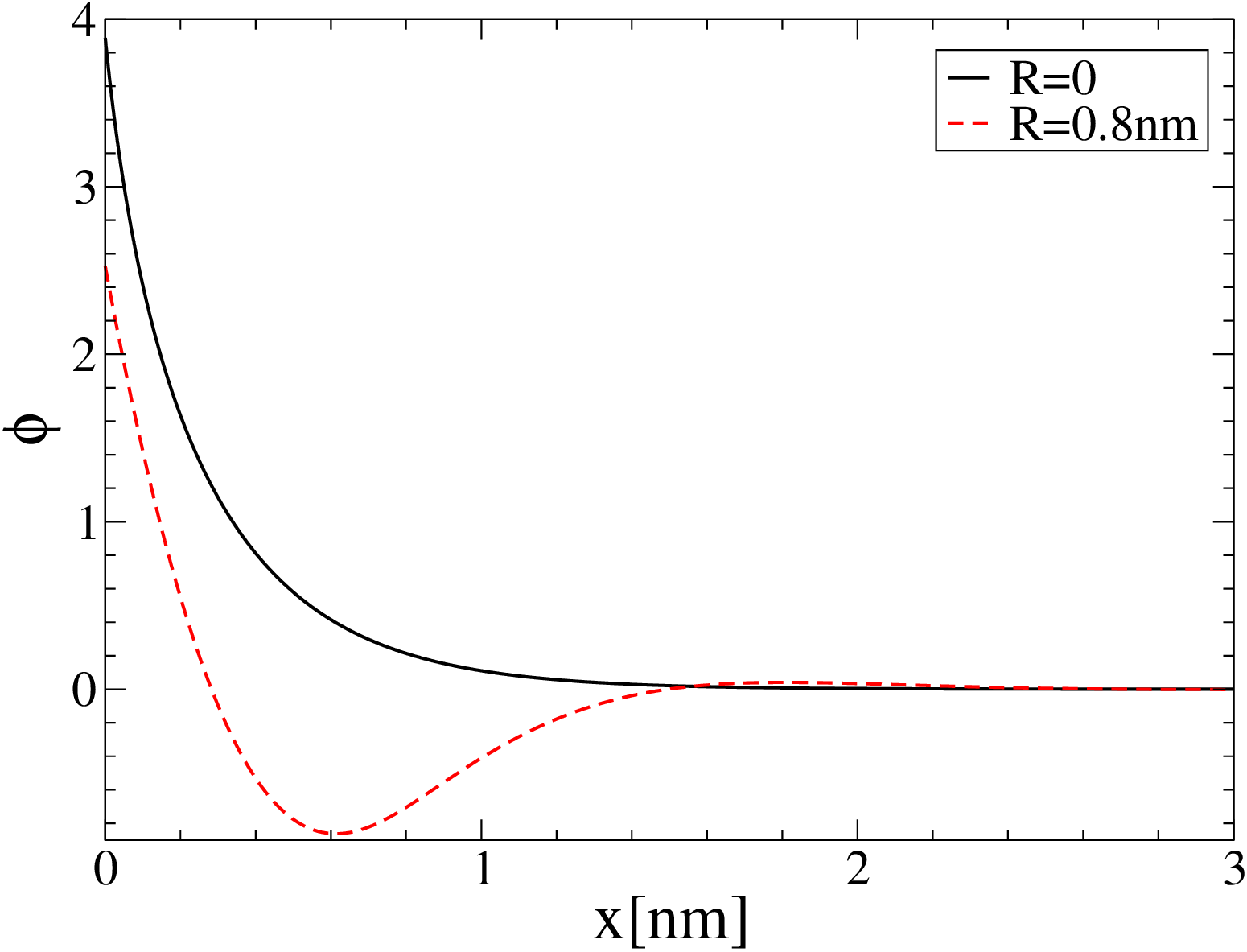}}}
\caption{Electrostatic potential near a charged wall corresponding to the system in
Fig.~(\ref{fig:rho_counter_R8_Z0_1.0M}).  The minimum corresponds to a point of vanishing 
electrostatic field.  }
\label{fig:phi_R8_Z0_1.0M}
\end{figure}

Fig.~(\ref{fig:rho_counter_R8_Z0_1.0M}) shows the counterion profiles for an electrolyte 
solution.  This introduces a length scale $\kappa^{-1}$.   Counterions from the bulk can now 
overcharge the surface.   The counterion profile beyond $x=0.5{\rm nm}$ dips below the bulk 
value, which indicates an overcharged surface.  Fig.~(\ref{fig:rho_co_R8_Z0_1.0M}) shows 
how the coion density rises above the bulk value,
another signature of charge reversal.  Finally, Fig. (\ref{fig:phi_R8_Z0_1.0M}) shows the 
non-monotonic electrostatic potential which goes to negative values and has a minimum around 
$x=0.6{\rm nm}$, at which point the electrostatic field vanishes and, farther on, it changes 
sign.  

Under what conditions penetrability of ions leads to charge reversal?  For penetration to take 
place, either the thermal fluctuations must exceed the interaction energy of the overlapped 
particles, $\beta U(r=0)\lesssim 1$, or the counterion concentration at the wall must be 
sufficiently large to involve numerous overlaps, such that the effective 2D density within the 
counterion layer is $\rho_{2D}R^2\gtrsim 1$.  For 
the distribution in Eq.~(\ref{eq:h1}) the first requirement translate into 
$R/\lambda_B\gtrsim 1.2$, and, assuming $\rho_{2D}\approx\sigma_c$, the second 
one becomes $\sigma_cR^2\gtrsim 1$.  
In  Fig.~(\ref{fig:rho_counter_R8_Z0_1.0M}) $R/\lambda_B\approx 1.1$ and 
$R^2\sigma_c\approx 1.6$, and so both criteria are met.

As a parenthetical note, we point out that the PB and the FSPB models satisfy the same 
contact value relation,
\begin{equation}
k_BT\rho_w = P_{\rm id} + \frac{\sigma_c^2}{2\epsilon},
\end{equation}
where $P_{\rm id}=k_BT\rho_b$ is the ideal gas pressure in the bulk, $\rho_b$ and $\rho_w$ is 
the total density in the bulk and at the contact with a wall, respectively.  

\subsection{Tuning of the short-range interactions}

The sole constraint that $\omega({\bf r}'-{\bf r})$ needs to satisfy is that its integral 
recovers the charge of an ion.  This leaves sufficient room to manipulate the short-range 
interactions.  As an example we can take a mixed distribution
\begin{eqnarray}
\omega_\pm(|{\bf r}-{\bf r}_i|) &=& 
\pm e\Bigg[({\mathcal Z}+1)\theta(R-|{\bf r}-{\bf r}_i|)/\nu\nonumber\\
&-&{\mathcal Z}\delta(R-|{\bf r}-{\bf r}_i|)/\gamma\Bigg],\nonumber\\
\label{eq:h2}
\end{eqnarray}
comprised of a charged shell and sphere.  $\gamma=4\pi R^2$ is the surface area of a sphere.  
${\mathcal Z}$ is the additional parameter and ${\mathcal Z}=0$ corresponds to the distribution 
in Eq. (\ref{eq:h1}).  The number density is
\begin{equation}
\rho_\pm = c_se^{\mp\big[({\mathcal Z}+1)\bar\phi_{\theta}-{\mathcal Z}\bar\phi_{\delta}\big]},
\end{equation}
where 
\begin{equation}
\bar\phi_{\delta}({\bf r}) = \frac{1}{\gamma}\int d{\bf r}' \phi({\bf r}')\delta(R-|{\bf r}'-{\bf r}|).
\end{equation}

\begin{figure}[tbh]
\vspace{0.6cm}
\centerline{\resizebox{0.45\textwidth}{!}
{\includegraphics{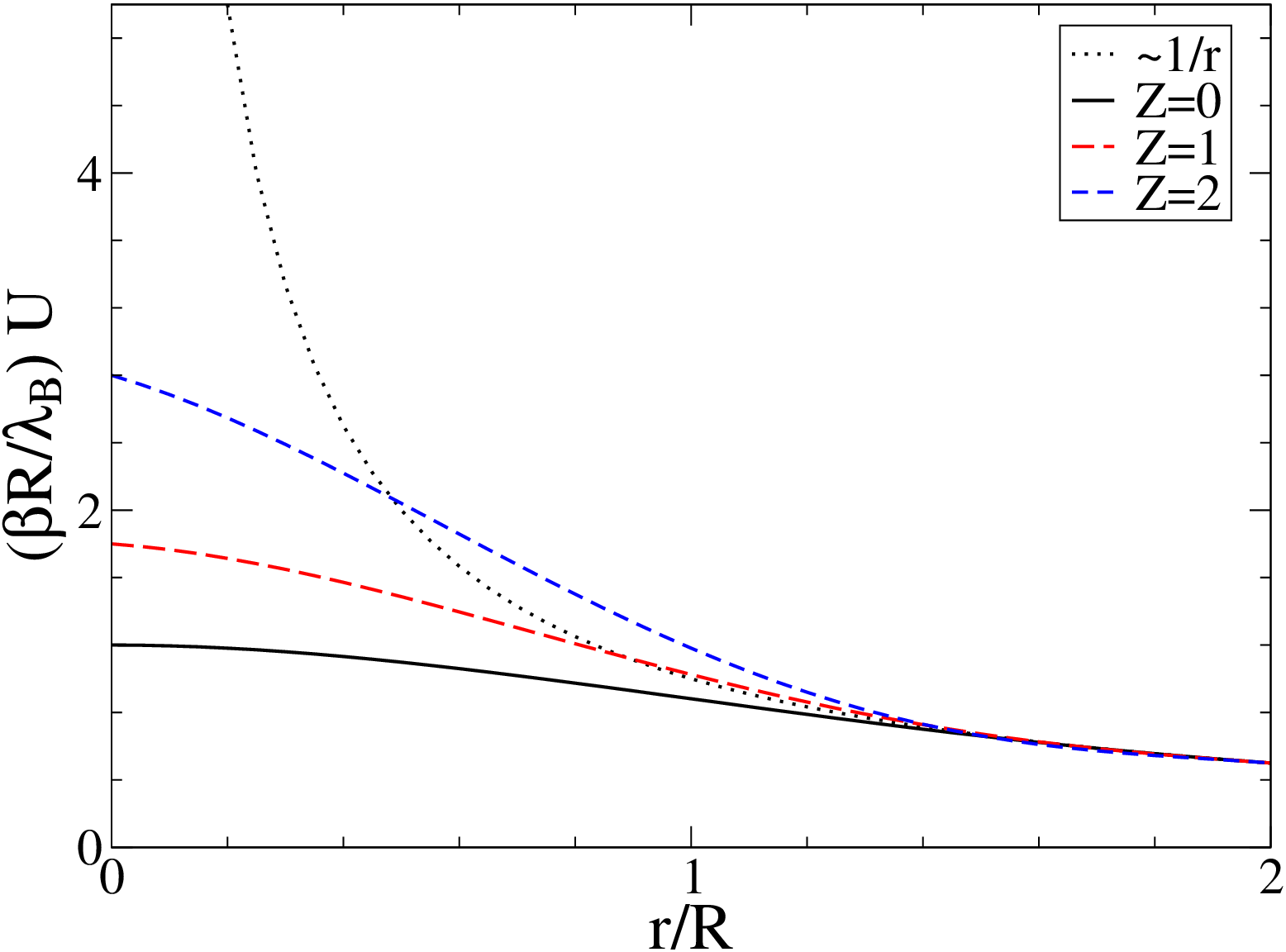}}}
\caption{Pair potential for two overlapping identical ions with charge distribution in 
Eq.~(\ref{eq:h2}).  ${\mathcal Z}=0$ corresponds to the distribution in Eq.~(\ref{eq:h1}).}  
\label{fig:U_Z}
\end{figure}

\begin{figure}[tbh]
\vspace{0.6cm}
\centerline{\resizebox{0.45\textwidth}{!}
{\includegraphics{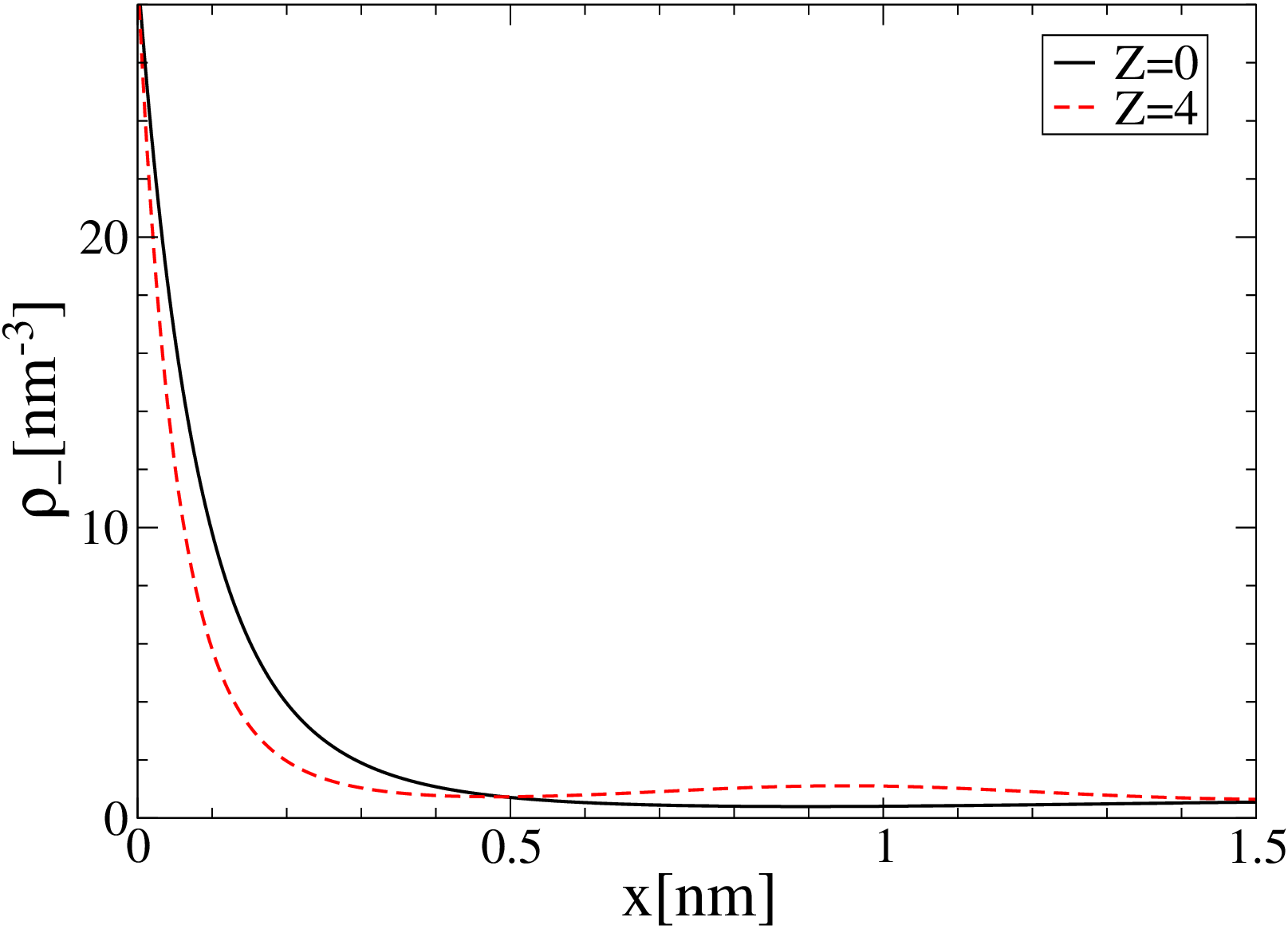}}}
\caption{The counterion density profile near a charged wall for the point ions and for the
the ions with the distribution in Eq. (\ref{eq:h2}) for $R=0.8{\rm nm}$ and different 
${\mathcal Z}$.  The system parameters are those in Fig.~(\ref{fig:rho_counter_R8_Z0_1.0M}). }
\label{fig:rho_counter_R8_Z4_1.0M}
\end{figure}

Fig.~(\ref{fig:U_Z}) plots the pair potentials for different ${\mathcal Z}$.  The parameter 
${\mathcal Z}$  controls the strength of the repulsion and, by the same token, the strength of 
the excluded volume interactions.   
By way of example, Fig.~(\ref{fig:rho_counter_R8_Z4_1.0M}) shows the density profile for 
${\mathcal Z}=0$ and ${\mathcal Z}=4$, for $R=0.8{\rm nm}$.  The excluded volume 
contributions for ${\mathcal Z}=4$ expel counterions from the first layer, which is the opposite of 
overcharging seen for ${\mathcal Z}=0$.  

With this example we try to demonstrate possible   
extensions of the model based on charge spreading.  For example, the repulsion controlled 
by the parameter ${\mathcal Z}$ could represent the interaction between polymers on account 
of the self-avoiding walk of polymer chains.  

\section{Comparison with Simulation}
To test the mean-field approximation implicit in the FSPB equation, we carry out Monte Carlo 
simulations.  The size of the simulation box is $L_y=L_z=8{\rm nm}$ and $L_x=6{\rm nm}$.  
The periodic boundary conditions are in the lateral $(y,z)$-directions.  The bounding plates at 
$x=0$ and $x=L_x$ have opposite charge.  The box encloses $N=600$ ions, $N_+=N_-=300$.  
To check for the finite size effects, we double the simulation size for some parameters.
The standard Ewald summation is applied for contributions of periodic images \cite{Frenkel02}.  
If particles overlap, we supplement the interaction energy with the term:  
$\beta\Delta U_{\rm tot}(r<2R)=\beta U(r_{ij})-\frac{\lambda_B}{r_{ij}}$.
(A physical picture is slightly modified when doing simulations.  In simulations we tend to think of 
particles as point charges which at separations $r<2R$ switch to the non-Coulomb interactions 
$U(r)$.  Within the mean-field theory based on the Poisson equation, 
we tend to think of an ion as a charged ball.  The two conceptions are, however, identical. )

\begin{figure}[tbh]
\vspace{0.6cm}
\centerline{\resizebox{0.45\textwidth}{!}
{\includegraphics{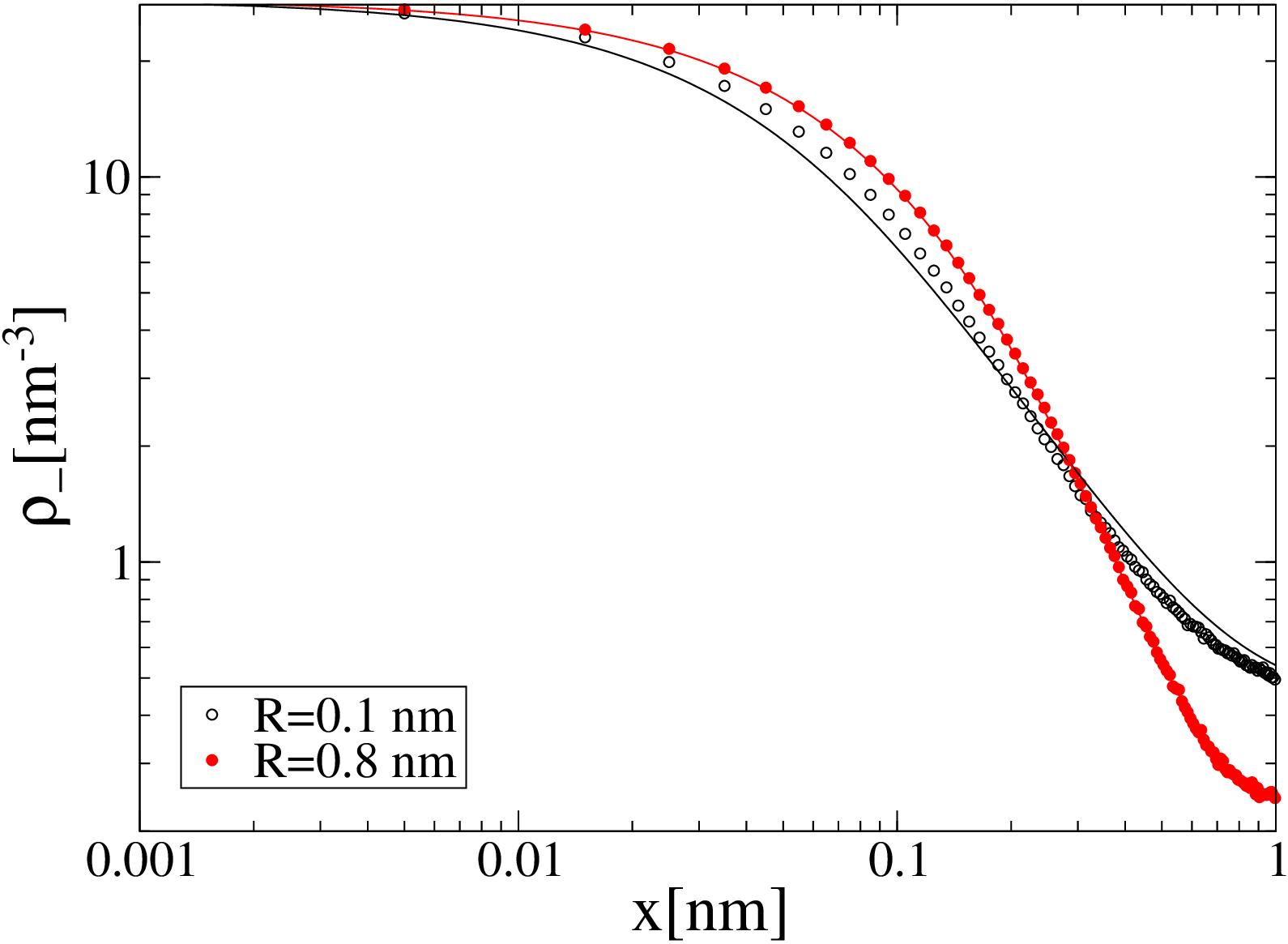}}}
\caption{Counterion density profiles.   
The system parameters are:   
$\lambda_{\rm GC}=0.09{\rm nm}$ and $\lambda_B=0.72{\rm nm}$.  The symbols designate  
the simulation data points and the lines are the numerical results for the FSPB equation.}
\label{fig:rho_counter_sim}
\end{figure}

\begin{figure}[tbh]
\vspace{0.6cm}
\centerline{\resizebox{0.45\textwidth}{!}
{\includegraphics{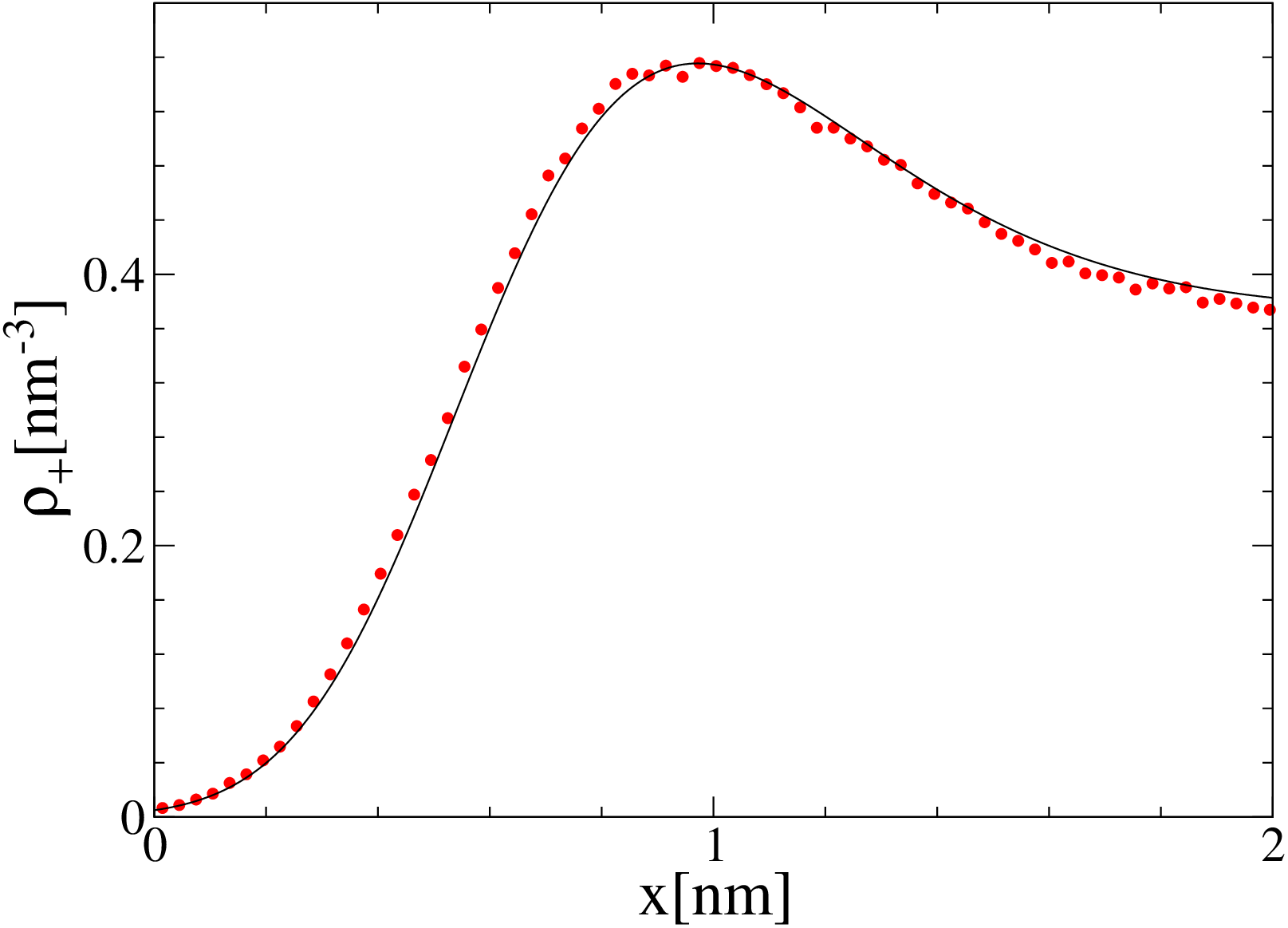}}}
\caption{The coion density profile for $R=0.8{\rm nm}$ obtained from the simulation (symbols) 
and the FSPB equation (solid line).  The remaining parameters are as in Fig.~(\ref{fig:rho_counter_sim}).}
\label{fig:rho_co_sim}
\end{figure}

Fig.~(\ref{fig:rho_counter_sim}) compares counterion profiles obtained from the simulation 
and the FSPB equation.  For $R=0.8{\rm nm}$ the agreement is virtually exact.  For 
the smaller radius $R=0.1{\rm nm}$ we see the two results deviate:  in the simulation, 
the correlations generate a more concentrated layer of counterions (but no overcharging).     
In Fig.~(\ref{fig:rho_co_sim}) we plot the coion density profile for $R=0.8{\rm nm}$ to
further confirm the accuracy of the mean-field in this regime.   
 
\begin{figure}[tbh]
\vspace{0.6cm}
\centering
\begin{tabular}{cc}
\includegraphics[width=40mm]{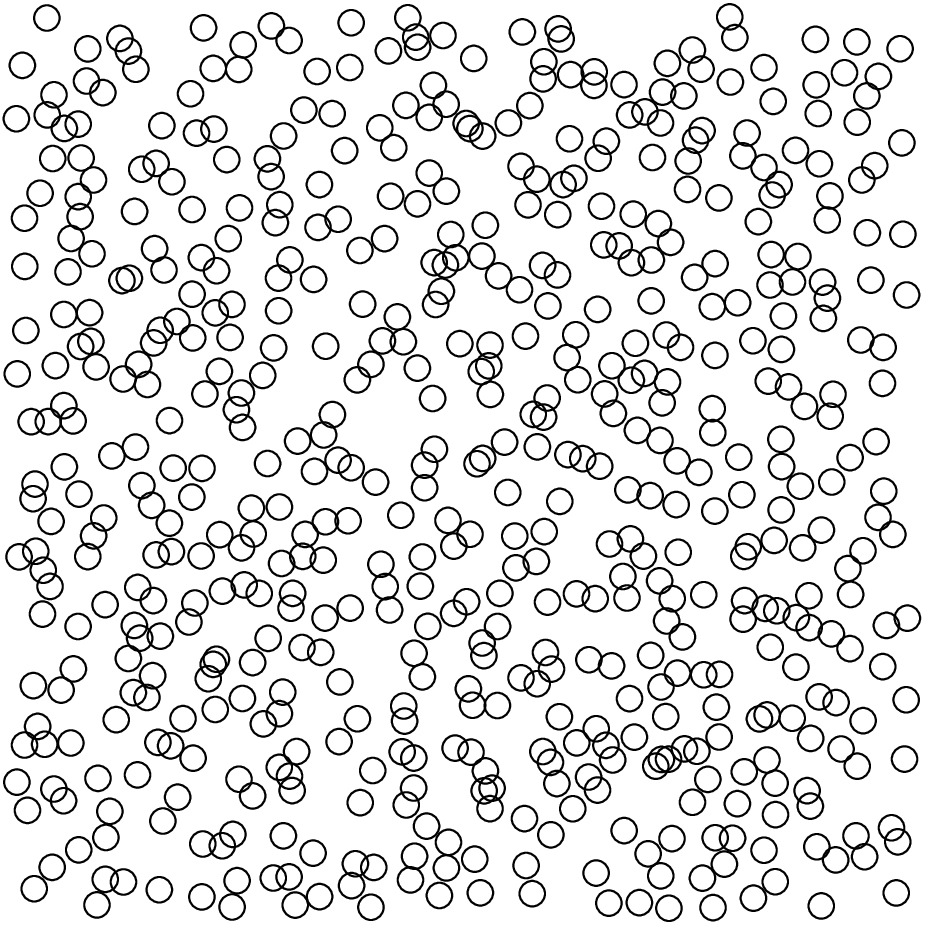}&
\hspace{0.4cm}
\includegraphics[width=40mm]{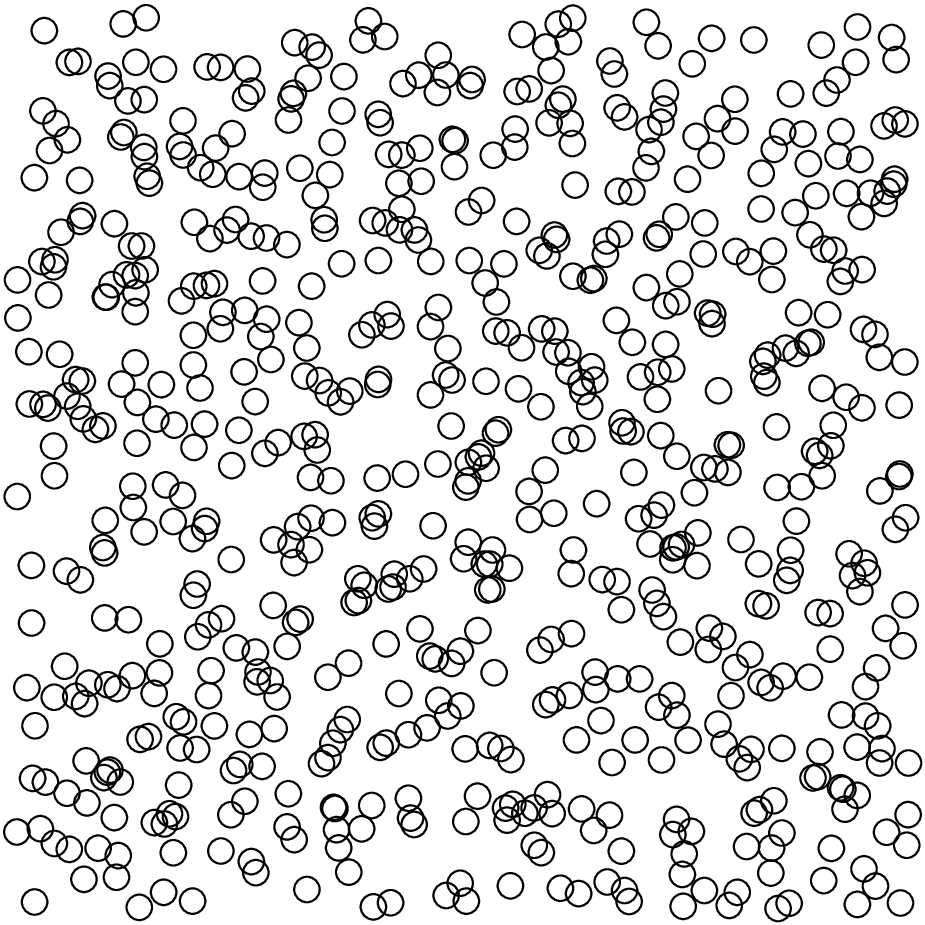}\\
\end{tabular}
\caption{Configuration snapshot of counterions adsorbed on the charged wall at 
$x<0.35{\rm nm}$.  The diameter of particles is $\sigma=0.5{\rm nm}$ and is selected arbitrarily 
for visualization.  The configuration on left is for $R/\lambda_B=0.14$ and the one on the right 
is for $R/\lambda_B=1.1$.  The 2d densities are $\rho_{2d}=2.34{\rm nm^{-2}}$ and 
$\rho_{2d}=2.68{\rm nm^{-2}}$, respectively.  For comparison, the surface charge density is 
$\sigma_c=2.50{\rm nm^{-2}}$.  The system parameters are as in 
Fig.~(\ref{fig:rho_counter_sim}). }
\label{fig:c1}
\end{figure}

What causes charge reversal for smeared-out ions?  Do counterions 
merely penetrate or the pair potential is sufficiently flat in the overlap region and counterions 
collapse into stacked formations, so that the pair correlation function exhibits positive peak 
as $r\to 0$?  In Fig.~(\ref{fig:c1}) we show the configuration snapshots for counterions 
adsorbed on a charged surface for different values of $R$.  The difference in structure is 
perceptible.  Ionic penetration favors smaller separations between counterions.  This gives 
the impression of string-like formations.  
As mentioned in the introduction, not all penetrable potentials can lead to stacked configurations,
where two or more particles occupy "the same" space.  For example, the Gaussian core model 
precludes \cite{Stillinger,Lowen00}, while the penetrable sphere model favors the stacked 
configurations \cite{Lowen98,Schmidt99}.  Stable stacked aggregates require 
that a pair potential be sufficiently flat.  A more rigorous test 
involves the Fourier transform of the pair interaction.   If $U(k)$ oscillates, and, therefore, 
involves negative values, then the stacking takes place under certain conditions \cite{Lowen01}.  
On the other hand, if $U(k)$ is non-negative, stacking does not occur under any conditions.  
This criterion is arrived at by considering the Ornstein-Zernike relation which, in the Fourier 
space, is
\begin{equation}
h(k) = \frac{c(k)}{1-\rho c(k)},
\end{equation}
where $c(r)$ and $h(r)$ is the direct and pair correlation function, respectively.  In the mean-field 
approximation $c_{\rm MF}(r)\approx -\beta U(r)$ [the exact definition is 
$c(r)=-\frac{\delta^2 \beta F_{ex}}{\delta\rho({\bf r})\delta\rho({\bf r}')}$, and in the 
mean-field $F_{ex}$ is the first term in Eq.~(\ref{eq:Omega})], so that
\begin{equation}
h_{\rm MF} = -\frac{\beta U(k)}{1+\rho\beta U(k)}.
\end{equation}
If we take $h(r=0)>0$ to be the signature of stacking, and 
\begin{equation}
h_{\rm MF}(0) = -\frac{1}{2\pi^2}\int_0^{\infty} dk\frac{\beta U(k)k^2}{1+\rho\beta U(k)},
\end{equation}
then it becomes immediately clear that a non-negative $U(k)$ always yields $h_{\rm MF}(0)<0$,
and no stacking occurs.   Only an oscillating $U(k)$ can yield $h_{\rm MF}>0$.  In the present 
model particles are smeared-out charges,
\begin{equation}
\beta U(|{\bf r}-{\bf r}'|) = \lambda_B\int d{\bf r}''\int{\bf r}'''\,
\frac{\omega({\bf r}-{\bf r}'')\omega({\bf r}'-{\bf r}''')}{|{\bf r}''-{\bf r}'''|},
\end{equation}
and in the Fourier space
\begin{equation}
\beta U(k)=\frac{4\pi\lambda_B\omega^2(k)}{k^2}.
\end{equation}
This yields
\begin{equation}
h_{\rm MF}(0) = -\frac{1}{2\pi^2}
\int_0^{\infty} dk\frac{4\pi\lambda_B\omega^2(k)k^2}{k^2+4\pi\lambda_B\rho\omega^2(k)}.
\label{eq:h_k}
\end{equation}
We see that $h(r=0)<0$ regardless of the distribution $\omega(r)$.  We conclude that soft 
interactions generated by charge spreading cannot lead to stacked configurations.    

\begin{figure}[tbh]
\vspace{0.6cm}
\centerline{\resizebox{0.45\textwidth}{!}
{\includegraphics{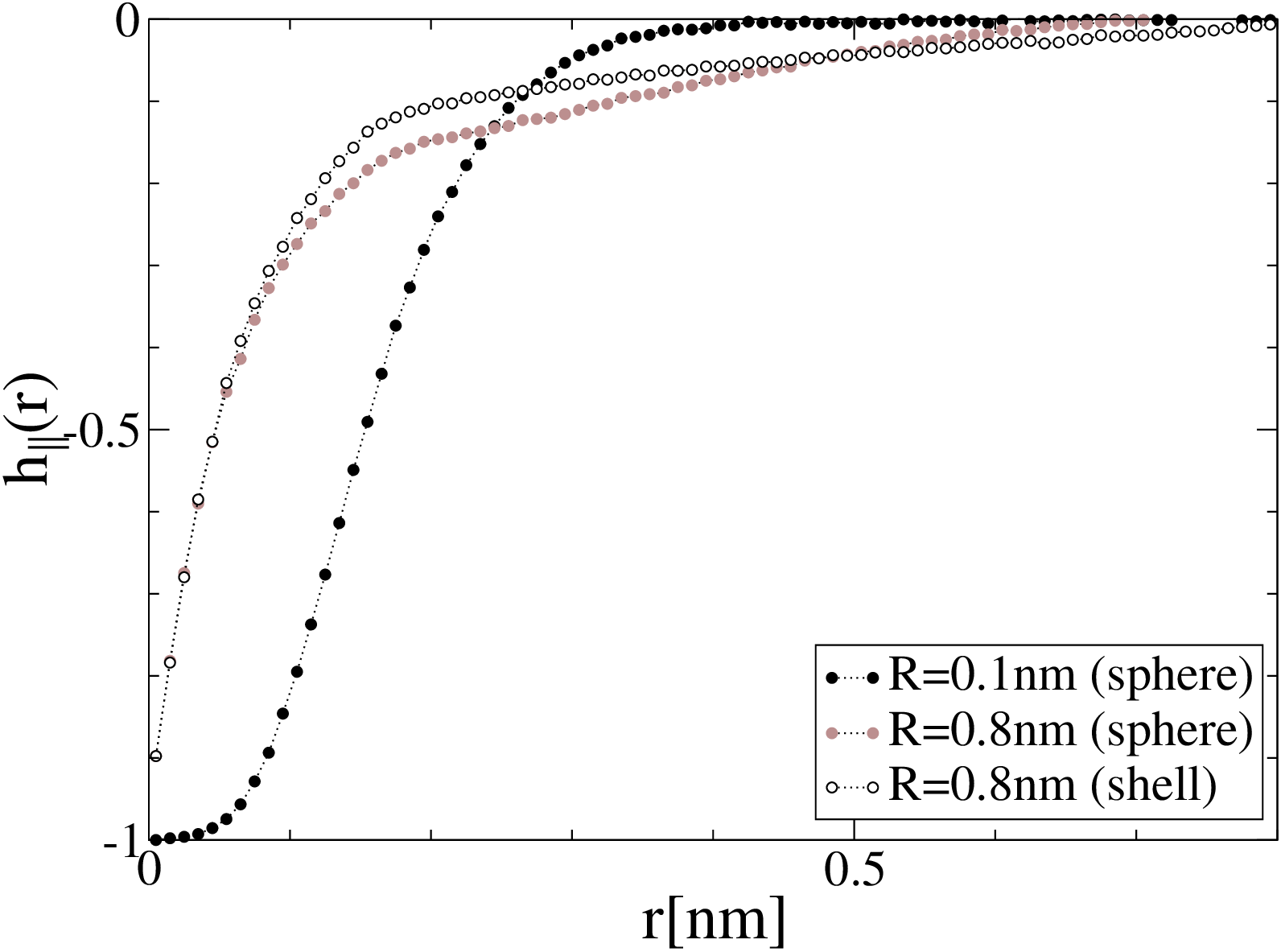}}}
\caption{Correlation function for counterions in the lateral plane adjacent 
to the wall.  The layer thickness is $0.35{\rm nm}$.  The remaining parameters are the same as in
Fig.~(\ref{fig:rho_counter_sim}).  The dashed lines guide the eye.  }
\label{fig:corr}
\end{figure}
\begin{figure}[tbh]
\vspace{0.6cm}
\centerline{\resizebox{0.45\textwidth}{!}
{\includegraphics{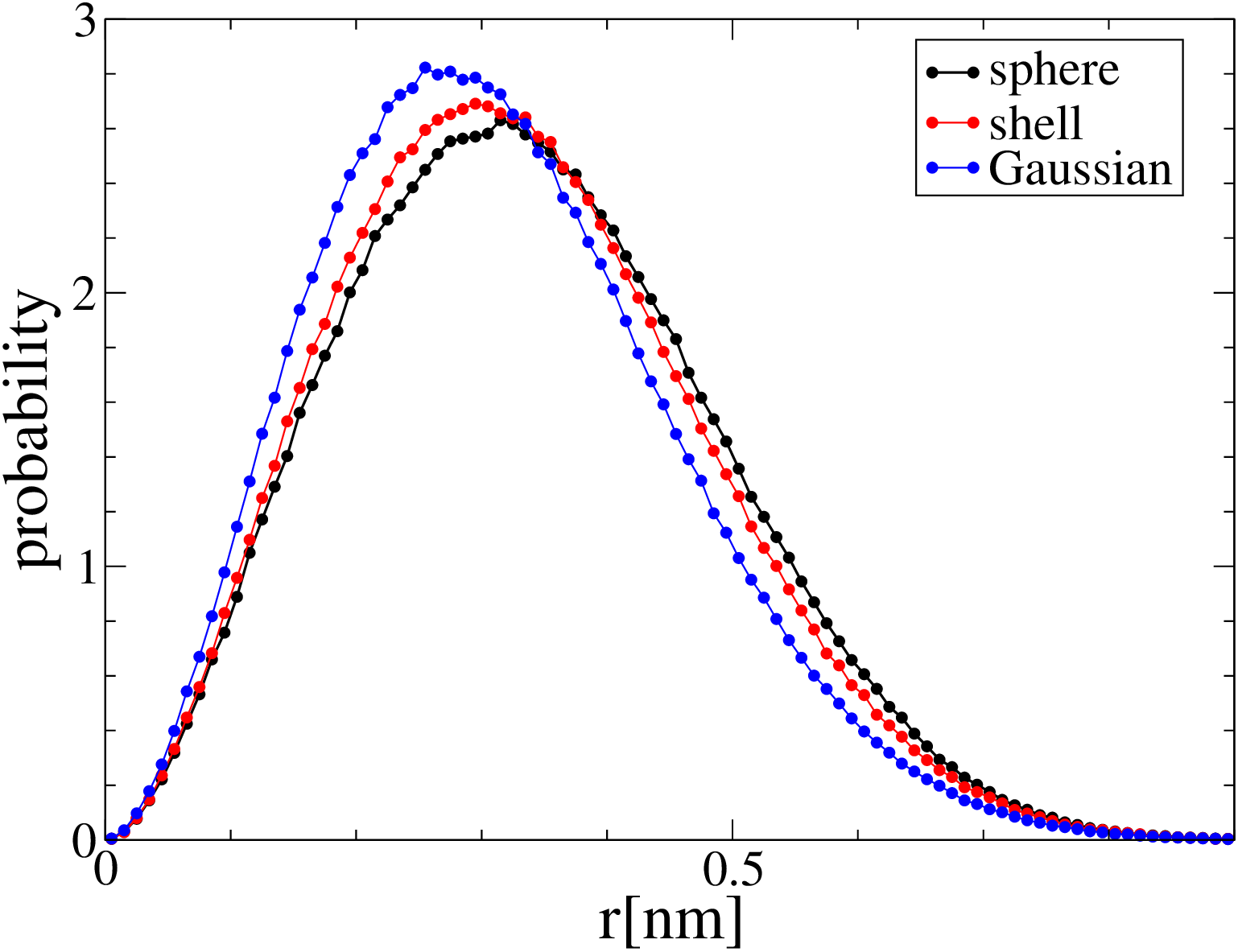}}}
\caption{The nearest neighbor separation distribution for different $\omega(r)$ functions.  
$R=0.8{\rm nm}$ and the same parameters as in Fig.~(\ref{fig:rho_counter_sim}).}
\label{fig:bond2}
\end{figure}

In Fig.~(\ref{fig:corr}) we show the lateral correlation function for the adsorbed coutnerions, 
$h_{\parallel}(r)$.  Penetration reduces the degree of electrostatic correlations between the 
ions, however, $h_{\parallel}(r)$ always remains negative and decreases monotonically all the 
way to $r=0$.   The small difference between the shell and the sphere distributions indicates that 
the exact shape of the pair potential is unimportant. 
In Fig.~(\ref{fig:bond2}) we show the probability distribution of the nearest neighbor separation 
for various $\omega(r)$ functions.  Differences are rather small and only quantitative.  The 
Gaussian $\omega(r)$ shows preference for smaller interionic separation.  

\begin{figure}[tbh]
\vspace{0.6cm}
\centerline{\resizebox{0.45\textwidth}{!}
{\includegraphics{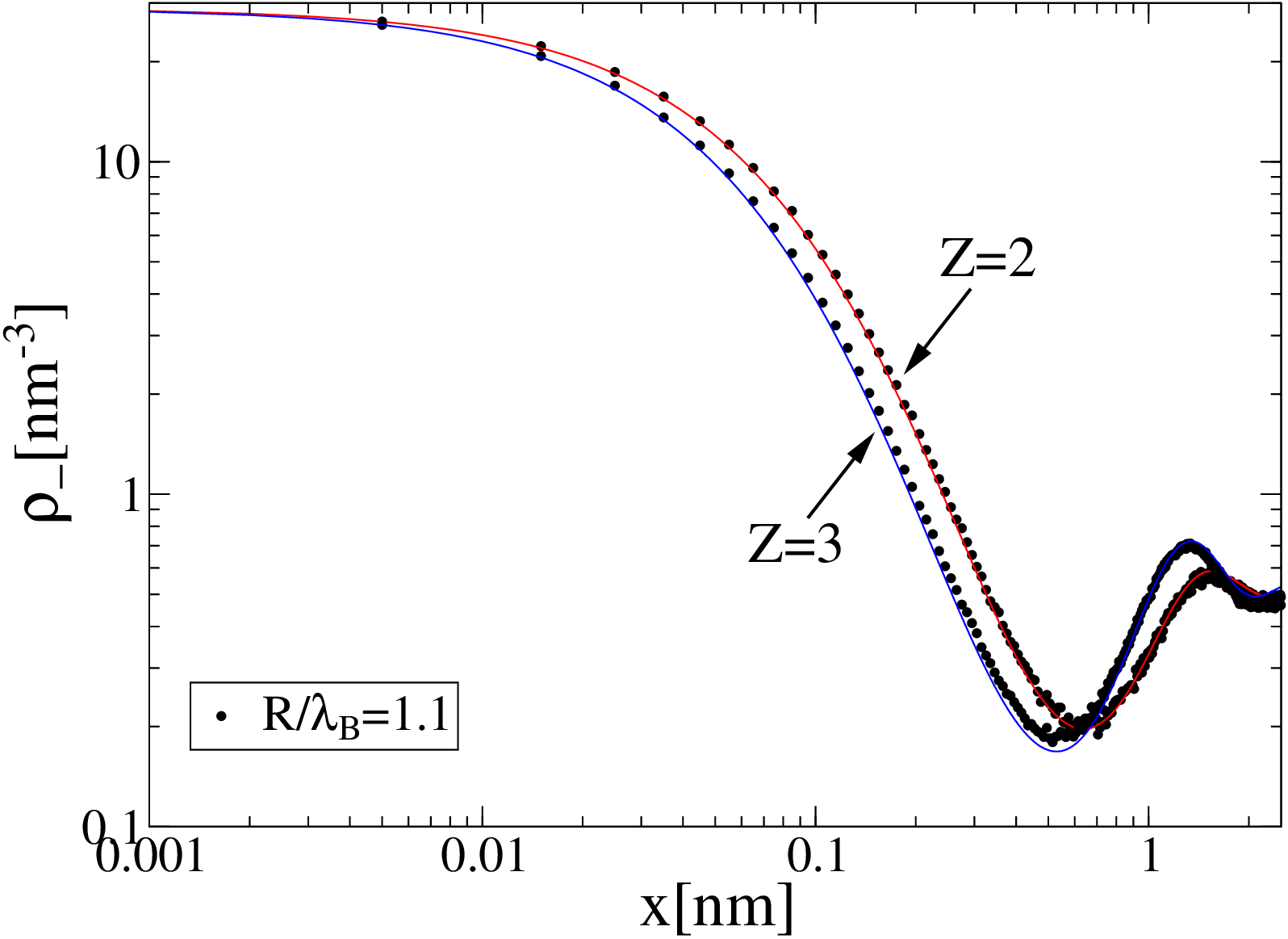}}}
\caption{Counterion distribution function for symmetric solution with ion charge $\pm eZ$. 
The remaining parameters are the same as in Fig.~(\ref{fig:rho_counter_sim}).  The circles
are the simulation data points and the lines are obtained from the FSPB. }
\label{fig:rho_counter_Q_sim}
\end{figure}

Finally, we explore the validity of the mean-field for multivalent symmetric ions, $Z>1$.  We 
suppose that the mean-field should deteriorate quickly for $Z>1$ since the coupling constant 
$\Gamma\sim Z^3$.  Fig.~(\ref{fig:rho_counter_Q_sim}) shows the counterion density 
profiles for $Z=2,3$.  For $Z=2$ the mean-field is virtually exact and for $Z=3$ there are small 
deviations.  This surprising agreement can be explained with the following.  Increasing $Z$ 
naturally reduces the number of counterions required to neutralize the wall.  On the other hand, 
a larger $Z$ generates stronger overcharging so the density drop is partially compensated.    
The number of overlapping configurations is still large and the mean-field retains its validity.  
To confirm this conjecture, we check the 2D 
density of the counterions adsorbed on the charged wall.  For $Z=3$, we find 
$R^2\rho_{2D}\approx 1.2 >1$.  The overlapping configurations are, therefore, still significant.  

\section{Discussion}
For overcharging to occur, there must exist a mechanism that either reduces the free energy of 
counterions condensed on a charged surface (it becomes favorable for a counterion to 
leave its bulk environment) or increase the free energy of a bulk (it becomes less favorable
for ions to stay in the bulk).  It is the exchange between the two environments, the bulk and the
double-layer, that leads to overcharging.  We consider the environment of a charged 
surface with its adsorbed counterions and consider the conditions that lead to lowering of its 
energy.  The surface charge of a plate is $\sigma_c$.  We do not violate the condition of 
neutrality, thus the 2D density of adsorbed counterions is $\sigma_c/Z$.  We focus on the 
energy of a single counterion.  For sake of simplicity, we assume counterions and the charged
wall occupy the same plane.  An 
attraction to the wall yields a negative energy contribution,
\begin{eqnarray}
\beta E_{\rm wall} &=& 
\lim_{r_c\to\infty}\Bigg[-2\pi\int_0^{r_c} dr\,r\frac{Z\lambda_B\sigma_c}{r}\Bigg] \nonumber\\
&=&\lim_{r_c\to\infty}\bigg[-2\pi Z\lambda_B\sigma_c r_c\bigg] 
\end{eqnarray}
which is countered by the repulsive interactions with other counterions,
\begin{equation}
\beta E_{\rm int} = \lim
_{r_c\to\infty}\Bigg[2\pi\int_0^{r_c} dr\,rg(r)\frac{Z\lambda_B\sigma_c}{r}\Bigg],
\end{equation}
where $g(r)$ is the pair distribution function. For absence of correlations $g(r)=1$, and 
the attractive contribution is completely canceled out, 
\begin{equation}
\beta E_{\rm int} = \lim_{r_c\to\infty}\bigg[2\pi Z\lambda_B\sigma_c r_c\bigg]. 
\end{equation}
But the cancellation will not be exact if correlations are present.   Separating correlations from the 
pair distribution function, $g(r)=1+h(r)$, the uncanceled correlation energy is
\begin{equation}
\beta E_{\rm corr} = 2\pi\lambda_B\sigma_c\int_{0}^{\infty} dr\,h(r),
\label{eq:E_h}
\end{equation}
The relation between the correlation energy and the correlation function is quite clear.  An ion
will generate a negative correlation hole in its neighborhood, which will lead to negative energy 
contributions.  We assume the following simple correlation hole, 
$h(r>\zeta)=0$ and $h(r<\zeta)=-1$, that is, by fixing a position of an ion, we introduce a 
circular hole in the density profile with the correlation length $\zeta$.  The conservation of mass 
condition requires that an area of the hole is related to the density via
$\sigma_c/Z=1/(\pi\zeta^2)$, therefore, the correlation length is $\zeta=\sqrt{Z/(\pi\sigma_c)}$.  
This correlation function is the low temperature limit of the correlation hole obtained by 
construction from the linear mean-field treatment \cite{Nordholm84}.  Inserting this $h(r)$ into 
Eq.~(\ref{eq:E_h}) we can approximate the 
correlation energy in the strong-coupling limit, 
\begin{equation}
\beta E_{\rm corr} \approx -2\pi^{1/2} Z^{3/2}\lambda_B\sigma_c^{1/2}.
\label{eq:correlation}
\end{equation}
Recalling the definition of the coupling constant $\Gamma=2\pi Z^3\lambda_B^2\sigma_c$,
we get
\begin{equation}
\beta E_{\rm corr} \approx -\sqrt{2\Gamma}.
\end{equation}

But for penetrable ions correlations are not required for reducing the electrostatic energy of 
counterions -- the fact demonstrated by the validity of the mean-field theory.  Consequently, 
we set $h(r)=0$.  The reduction in electrostatic energy comes from a different source, from the 
fact that at overlapping separations, $r<2R$, the electrostatic interactions are reduced on 
account of smearing procedure, which leads to the energy gain
\begin{eqnarray}
\beta E_{\rm overlap} &\approx& 
2\pi\int_0^{2R} dr\, \sigma_c r\Bigg[\beta U(r)-\frac{Z\lambda_B}{r}\Bigg]\nonumber\\
&=& -C\pi Z\lambda_B\sigma_c R,
\label{eq:overlap}
\end{eqnarray}
where the constant $C$ depends on the pair potential $U(r)$, which, in turn, is determined by 
the distribution function $\omega(r)$.  For the distributions considered in this work:  
$C_{\rm sphere}=36/35$, $C_{\rm shell}=4/3$, and $C_{\rm gauss}=2.06$.  
Comparing Eq.~(\ref{eq:correlation}) with Eq.~(\ref{eq:overlap}) we see the different 
dependence of each mechanism on different parameters.  The stabilization based on 
penetration has stronger dependence on the surface charge and the Bjerrum constant, 
on the other hand, its dependence on valency is weaker.  
In the strong-coupling limit, penetrable ions will exhibit correlated motions.  
If the correlation length is larger than the diameter of an ion, $\sqrt{Z/(\pi\sigma_c)}>2R$, 
penetration may be neglected and the former mechanism comes to the fore.  On the other hand, 
if $\sqrt{Z/(\pi\sigma_c)}<2R$, we expect the two contributions to mix.

To recap, both mechanisms depend on eliminating the energy contributions coming from short
separations between ions.  For point ions in the strong-coupling limit, configurations with short 
separations are eliminated through correlated motion.  The price is sacrifice in entropy, despite 
this, the total free energy is lowered.  
For penetrable ions the problem of high energy contributions at short separations does not exist
to begin with.  Due to smearing out procedure of an ion charge and the removal of the divergence 
from pair interactions, these contributions are taken out of the picture.  
Consequently, there is no entropy price 
to be paid, as all separations are explored "equally" and the mechanism is valid in the 
weak-coupling limit.


\section{Conclusion}
The present work studies the structure of a double-layer composed of ions whose central charge is 
smeared  over a finite region in accordance with a weight function $\omega(r)$.  
The smearing-out procedure removes the divergence as $r\to 0$ from the pair interaction, 
allowing for interpenetration between the ions.  The conditions under which
penetration is favored are large temperature and high density.  This regime is suitable for the 
mean-field treatment.  Accordingly, we derive a modified Poisson-Boltzmann equation for
spread-out charges (the FSPB equation).  The FSPB equation predicts that for  
sufficiently large spreading radius $R$, overcharging takes place  -- the MC simulations verify
this prediction.  This suggests an alternative 
mechanism for charge reversal that is not related to correlations and the strong-coupling limit.  

Using simulations and the mean-field approximation, we can exclude the ionic stacking as the 
underlying mechanism of overcharging.  In fact, any soft repulsion obtained by charge spreading
cannot lead to stacked configurations.  


As a final consideration, we address the physical relevance of the smeared-out ion model.  
The spreading-out of the charge may capture the interactions between charges distributed 
along the polymer chains, but a more realistic representation would involve a non-electrostatic 
component produced by the self-avoidance of the polymer chains.  
Within the electrostatics framework, we have suggested a plausible distribution
$\omega(r)$ composed of a charged shell and sphere, which generates an 
additional repulsion inside a penetrable core.  In the end, however, a physically accurate pair 
potential requires corroboration with experimental analysis.

\begin{acknowledgments}
D.F. is grateful to Nathan Berkovits for an invitation to and short sojourn in the ICTP-SAIFR 
where, prompted by the theoretical atmosphere of the institute, the formal aspects of 
the FSPB equation were laid down. This work was partially supported by the CNPq, FAPERGS, INCT-FCx, and by the US-AFOSR under the grant FA9550-12-1-0438.
\end{acknowledgments}



\end{document}